\documentstyle[12pt]{article}
\begin{document}

\tolerance=5000

\def\pp{{\, \mid \hskip -1.5mm =}}
\def\cL{{\cal L}}
\def\be{\begin{equation}}
\def\ee{\end{equation}}
\def\bea{\begin{eqnarray}}
\def\eea{\end{eqnarray}}
\def\tr{{\rm tr}\, }
\def\nn{\nonumber \\}
\def\e{{\rm e}}

\ 

\vskip -2cm

\ \hfill
\begin{minipage}{3.5cm}
NDA-FP-44 \\
February 1997 \\
\end{minipage}

\vfill

\begin{center}
{\Large\bf Effective Action for Conformal Scalars and 
Anti-Evaporation of Black Holes}

\vfill

{\large\sc Shin'ichi NOJIRI}\footnote{e-mail : nojiri@cc.nda.ac.jp} 
and {\large Sergei D. ODINTSOV$^\clubsuit$}\footnote{e-mail : 
odintsov@galois.univalle.edu.co, 
odintsov@kakuri2-pc.phys.sci.hiroshima-u.ac.jp}

\vfill

{\large\sl Department of Mathematics and Physics \\
National Defence Academy, Hashirimizu Yokosuka 239, JAPAN}

{\large\sl $\clubsuit$ Tomsk Pedagogical University, 
634041 Tomsk, RUSSIA \\
and \\
Dep.de Fisica, Universidad del Valle \\
AA25360, Cali, COLOMBIA \\
}

\vfill

{\bf abstract}

\end{center}

We study the one-loop effective action for $N$ 4D conformally 
invariant scalars on the spherically symmetric background. 
The main part of effective action is given by integration of 
4D conformal anomaly. This 
effective action (in large $N$ approximation and partial curvature 
expansion) is applied to investigate the quantum evolution of 
Schwarzschild-de Sitter 
(SdS) black holes of maximal mass. We find that the effect 
(recently discovered 
by Bousso and Hawking for $N$ minimal scalars and another 
approximate 
effective action) of anti-evaporation of nearly maximal SdS (Nariai)
 black holes takes also place in the model under consideration. 
Careful treatment of quantum corrections and perturbations modes 
of Nariai black hole is given being quite complicated. 
It is shown that exists also 
perturbation where black hole radius shrinks, i.e. black hole 
evaporates. 
We point out that our result holds for wide class of models 
including conformal scalars, spinors and vectors. 
Hence,anti-evaporation of SdS black holes 
is rather general effect which should be taken into account in 
quantum gravity considerations.

\ 


\newpage

\section{Introduction.}

The gravitational black holes radiation \cite{Ha} may be considered as the one of 
the most brilliant manifestations of quantum gravity. Moreover, one can expect 
that other quantum gravity strong effects should be also searched in black holes.
Recently, one such new effect (anti-evaporation of black holes) has been found 
in ref.\cite{BH}. The authors of above work used large $N$ and $s$-waves approximations 
in their calculation of effective action for 4D minimal scalar. This effective action 
has been used to study the quantum stability of the Schwarzschild-de Sitter black 
holes of maximal mass (Nariai solution). It has been shown that there is the perturbation 
mode where the black hole size increases and the black hole grows back to the maximal 
radius (anti-evaporation).

The interesting question is if that effect is quite common or it results from the 
specific model and (or) approximation used. To answer this question we study another model 
(of 4D conformal scalars) 
and find the effective action in another approximation (large $N$ and expansion on the 
curvature). The analysis of quantum corrected equations of motion is much more complicated 
than in Bousso-Hawking model. Nevertheless, the final result is quilitatively the same-
we confirm the possibility of black holes anti-evaporation.

The contents of the paper are following. In the next section 4D anomaly induced effective action 
for conformal scalar is evaluated. We work in large $N$ approximation and calculate the piece of
the effective action which is not anomaly induced as expansion on the curvature near spherically 
symmetric background. As a result we obtain effectively 2D gravitational system. Equations of motion are derived 
in the section 3. Section 4 is devoted to the analysis of quantum perturbations for Nariai 
black hole. We demonstrate the possibility of anti-evaporation of such objects. In the last section 
we give some comments and explain that our result is of general relevancy for a class of quantum conformal matter 
models.

\section{Effective action for conformal scalars}

In the present section we derive the effective action for 
conformally invariant scalars (for a general review of 
effective action in curved space, see\cite{BOS}).
Let us start from Einstein gravity with $N$ 
conformal scalars $\chi_i$:
\bea
\label{OI}
S&=&-{1 \over 16\pi G}\int d^4x \sqrt{-g_{(4)}}
\left\{R^{(4)} -2\Lambda\right\} \nn
&& + {1 \over 2}\sum_{i=1}^N \int d^4x \sqrt{-g_{(4)}}
\left(g_{(4)}^{\alpha\beta}\partial_\alpha\chi_i
\partial_\beta\chi_i + {1 \over 6}R^{(4)}\chi_i^2 \right)\ .
\eea
The convenient choice for the spherically symmetric 
spacetime is the following:
\be
\label{OII}
ds^2=g_{\mu\nu}dx^\mu dx^\nu + f(\phi) d\Omega
\ee
where $\mu,\nu=0,1$, $g_{\mu\nu}$ and $f(\phi)$ depend only 
from $x^0$, $x^1$.

Reducing 4D action (\ref{OI}) in accordance with the metric 
(\ref{OII}) and working in $s$-wave sector (i.e., 
$\chi_i=\chi_i(x^0,x^1)$), we get
\bea
\label{OIII}
{S \over 4\pi}&=&-{1 \over 16\pi G}\int d^2x \sqrt{-g}
\left[ f(\phi) (R - 2\Lambda) + R_\Omega 
+2 (\nabla^\mu f^{1 \over 2})(\nabla_\mu f^{1 \over 2})
\right] \nn
&&+ {1 \over 2}\sum_{i=1}^N \int d^2x \sqrt{-g}
\left[f(\phi)\nabla^\alpha\chi_i
\nabla_\alpha\chi_i + {1 \over 6}\chi_i^2(Rf(\phi)
+ R_\Omega)\right] .
\eea
where factor $4\pi$ appears as volume of $S^2$ space 
and $R_\Omega=2$ (or $R_\Omega={2 \over \rho^2}$ 
where $\rho^2=1$). If we take $R_\Omega={2 \over \rho^2}$, 
then $V_{S_2}=4\pi\rho^2$.

Let us start the calculation of effective action due to 
scalars on the background (\ref{OII}). 4D scalar in 
Eq.(\ref{OI}) is conformally invariant. Let us rewrite the 
metric (\ref{OII}) as following:
\be
\label{OIV}
ds^2=f(\phi)\left[
f^{-1}(\phi)g_{\mu\nu}dx^\mu dx^\nu + d\Omega\right]\ .
\ee
In the calculation of effective action, we present 
effective action as following:
\be
\label{OV}
\Gamma=\Gamma_{ind}+ \Gamma[1, g^{(4)}_{\mu\nu}]
\ee
where $\Gamma_{ind}=\Gamma[f, g^{(4)}_{\mu\nu}]
- \Gamma[1, g^{(4)}_{\mu\nu}]$ is conformal anomaly 
induced action which is quite well-known \cite{R}, 
$g^{(4)}_{\mu\nu}$ is metric (\ref{OIV}) without 
multiplier in front of it, i.e., $g^{(4)}_{\mu\nu}$ 
corresponds to 
\be
\label{OVI}
ds^2=\left[
f^{-1}(\phi)g_{\mu\nu}dx^\mu dx^\nu + d\Omega\right]\ .
\ee

The conformal anomaly for $N$ 4D scalars is well-known 
one:
\be
\label{OVII}
T=b\left(F+{2 \over 3}\Box R\right) + b' G + b''\Box R
\ee
where $b={N \over 120(4\pi)^2}$, 
$b'=-{N \over 360(4\pi)^2}$, $b''=0$ but 
in principle, $b''$  may be changed by the finite 
renormalization of local counterterm, $F$
is the square of Weyl tensor, $G$ is 
Gauss-Bonnet invariant.

Conformal anomaly induced effective action $\Gamma_{ind}$ 
may be written as follows \cite{R}:
\bea
\label{OVIII}
W&=&b\int d^4x \sqrt{-g} F\sigma \nn
&& +b'\int d^4x \sqrt{-g} \Bigl\{\sigma\left[
2\Box^2 + 4 R^{\mu\nu}\nabla_\mu\nabla_\nu 
- {4 \over 3}R\Box + {2 \over 3}(\nabla^\mu R)\nabla_\mu 
\right]\sigma \nn
&& + \left(G-{2 \over 3}\Box R\right)\sigma \Bigr\} \nn
&& -{1 \over 12}\left(b'' + {2 \over 3}(b + b')\right)
\int d^4x \sqrt{-g}\left[R - 6 \Box \sigma 
- 6(\nabla \sigma)(\nabla \sigma) \right]^2 
\eea
where $\sigma={1 \over 2}\ln f(\phi)$, and 
$\sigma$-independent terms are dropped. All 4-dimensional 
quantities (curvatures, covariant derivatives) in 
Eq.(\ref{OVIII}) should be calculated on the metric 
(\ref{OVI}). (We did not write subscript $(4)$ for them.)
Note that after calculation of (\ref{OVIII}) on the metric 
(\ref{OVI}), we will get effectively two-dimensional 
gravitational theory.

The next step is to calculate second term in the right 
hand side of Eq.(\ref{OV}), i.e., conformally invariant part of EA.
We work in $s$-wave approximation where scalars depend only 
from first two coordinates. So we have to make 
reduction for scalar action on the metric (\ref{OVI}). 
We present now the metric (\ref{OVI}) as follows:
\be
\label{OIX}
ds^2=\tilde g_{\mu\nu}dx^\mu dx^\nu + d\Omega
\ee
where $\tilde g_{\mu\nu}=f^{-1}(\phi) g_{\mu\nu}$.

On the space with the background (\ref{OIX}), having in 
mind $s$-wave approximation and reduction, we get the 
following two dimensional classical action for scalars:
\be
\label{OX}
{S \over 4\pi}= {1 \over 2}\sum_{i=1}^N \int d^2x 
\sqrt{-\tilde g}
\Biggl[\tilde\nabla^\alpha\chi_i\tilde\nabla_\alpha\chi_i 
+ {1 \over 6}\chi_i^2\left(\tilde R + R_\Omega) 
\right)\Biggr] .
\ee
where all quantities in (\ref{OIX}) (i.e., $\tilde R$, 
$\tilde{\tilde\Box}$ and 
$\tilde\nabla_\alpha$) should be calculated on 
two-dimensional metric $\tilde g_{\mu\nu}$. As one can
easily see second term in (\ref{OX}) breaks the 
conformal invariance. 
Denote 
${\tilde R + R_\Omega \over 6}\equiv V$.

It is clear that it is impossible to calculate the finite 
part of effective action corresponding to quantum system 
(\ref{OX}) in closed form. So, we will develop expansion on 
the ``potential term'' $V$. Zero term of this expansion 
(i.e., when we put $V=0$) may be found in closed form 
as it is standard Polyakov anomaly induced action \cite{P}:
\be
\label{OXI}
\Gamma_{(0)}={N \over 96\pi}\int d^2x \sqrt{-\tilde g}
\tilde R {1 \over \tilde{\bar \Box}} \tilde R\ .
\ee
The first order term on the potential $V$ may be found 
using zeta-regularization technique 
(see, \cite{E}
 for a review). Actually, this term 
represents Coleman-Weinberg type term:
\be
\label{OXII}
\Gamma_{(V)}={N \over 8\pi}\int d^2x \sqrt{-\tilde g}
\left( V - V \ln{V \over \mu^2} \right)
\ee
where $\mu^2$ is an arbitary mass parameter.
Note that one can calculate the next terms of the 
expansion on $V$ using different techniques. 
If we suppose that curvature $\tilde R$ is quickly changing 
than next leading term of expansion is of the form 
(\ref{OXI}) with another (significantlly smaller) 
coefficient than in (\ref{OXI}). 

Finally effective action in our model will take the 
following form:
\be
\label{OXIII}
\Gamma = \Gamma_{ind}-\Gamma_{(0)} - 
\Gamma_{(V)}
\ee
where $\Gamma_{ind}$ is given by Eq.(\ref{OVIII}) and all 
quantities in Eq.(\ref{OVIII}) should be calculated on metric 
(\ref{OVI}), $\Gamma_{(0)}$ is given by two-dimensional 
expression (\ref{OXI}) and $\Gamma_{(V)}$ is given by 
Eq.(\ref{OXII}). The quantum effective action (\ref{OXIII}) 
should be added to classical action (\ref{OIII}). That is 
combined action $S+\Gamma$ which will define the 
quantum dynamics of the system under discussion.

%%%%%%%%%%%

%%%%%%%%%%%%%%%%%%

We now consider to solve the equations of motion obtained from 
the above effective Lagrangians $S+\Gamma$. 
In the following, we use $\tilde g_{\mu\nu}$ and $\sigma$ 
as a set of independent variables and we write 
$\tilde g_{\mu\nu}$ as $g_{\mu\nu}$ if there is not 
confusion.
First we should note that 
the definitions of the Gauss-Bonnet invariant $G$ and 
the Weyl tensor $C_{\mu\nu\alpha\beta}$ are given by
\bea
\label{GBinv}
G&\equiv& R_{\mu\nu\alpha\beta}R^{\mu\nu\alpha\beta}
-4 R_{\alpha\beta} R^{\alpha\beta} + R^2 \\
\label{Wyl}
C_{\mu\nu\alpha\beta}&\equiv&R_{\mu\nu\alpha\beta}
+ {1 \over 2}(g_{\mu\beta}R_{\nu\alpha}
-g_{\mu\alpha}R_{\nu\beta}+g_{\nu\alpha}R_{\mu\beta}
-g_{\nu\beta}R_{\mu\alpha}) \nn
&& + {1 \over 6}R(g_{\mu\alpha}g_{\nu\beta} 
- g_{\mu\beta}g_{\nu\alpha}) \ .
\eea
Then the square of the Weyl tensor has the following form:
\be
\label{Wylsq}
C_{\mu\nu\alpha\beta}C^{\mu\nu\alpha\beta}
= R_{\mu\nu\alpha\beta}R^{\mu\nu\alpha\beta}
-2 R_{\alpha\beta} R^{\alpha\beta} + {1 \over 3}R^2 
\ee
Then $\Gamma_{ind}$ ($W$ in Eq.(\ref{OVIII})) 
is rewritten after the reduction to 2 dimensions as follows:
\bea
\label{Gindrd}
{\Gamma_{ind} \over 4\pi}&=&
b\int d^2x\sqrt{-g}\left(
 R^{(2)}_{\mu\nu\alpha\beta}R^{(2)\mu\nu\alpha\beta}
-2 R^{(2)}_{\alpha\beta} R^{(2)\alpha\beta} 
+ {1 \over 3}(R^{(2)})^2 \right. \nn
&& \left. + {2 \over 3}R_\Omega R^{(2)} 
+ {1 \over 3}R_{\Omega}^2 
\right) \sigma \nn
&& + b'\int d^2x\sqrt{-g}\left\{\sigma \left(
2\Box^2 + 4R^{(2)\mu\nu}\nabla_\mu \nabla_\nu 
-{4 \over 3}(R^{(2)} + R_\Omega)\Box \right.\right. \nn
&& \left. 
+ {2 \over 3}(\nabla^\mu R^{(2)})\nabla_\mu\right) 
\sigma \nn
&& + \left(R^{(2)}_{\mu\nu\alpha\beta}
R^{(2)\mu\nu\alpha\beta}
-4 R^{(2)}_{\alpha\beta} R^{(2)\alpha\beta} + (R^{(2)})^2 
\right. \nn 
&& \left. \left. 2R_\Omega R^{(2)} - {2 \over 3}\Box R^{(2)}
\right)\sigma \right\} \nn
&& -{1 \over 12}\left\{ b'' + {2 \over 3}(b+b')\right\}
\int d^2x\sqrt{-g} \nn
&& \times \left\{ \left( R^{(2)} + R_\Omega - 6 \Box\sigma 
- 6\nabla^\mu\sigma\nabla_\mu\sigma \right)^2 
- \left( R^{(2)} + R_\Omega \right)^2 \right\}\ .
\eea
Here $R_\Omega=2$ is scalar curvature of $S^2$ with 
the unit radius. The suffix ``(2)'' expresses the quantitiy in 2 
dimensions but we abbreviate it if there is no any confusion.
We also note that in two dimensions the Riemann tensor $R_{\mu\nu\sigma\rho}$ 
and $R_{\mu\nu}$ are expressed via the scalar curvature $R$ and the metric 
tensor $g_{\mu\nu}$ as follows
\be
\label{Rs2d}
R_{\mu\nu\sigma\rho}={1 \over 2}\left(g_{\mu\sigma}g_{\nu\rho}
- g_{\mu\rho}g_{\nu\sigma}\right)R\ ,\ \ \ R_{\mu\nu}={1 \over 2}g_{\mu\nu}R\ .
\ee

\section {Equations of motion}

Let us derive the equations of motion 
with account of quantum corrections from above effective action.
In the following, we work in  the conformal gauge
\be
\label{cf}
g_{\pm\mp}=-{1 \over 2}\e^{2\rho}\ ,\ \ \ g_{\pm\pm}=0
\ee
after considering the variation of the effective action 
$\Gamma+S$ with respect to $g_{\mu\nu}$ and $\sigma$.
Note that the tensor $g_{\mu\nu}$ under consideration is the product of the 
original metric tensor in (\ref{OIX}) and the $\sigma$-function 
$\e^{-2\sigma}$, the equations given by the variations of 
$g_{\mu\nu}$ are the combinations of the equations given by the 
variation of the original metric and $\sigma$-equation.

Since $g^{++}$ and $g^{--}$ vanish in the conformal gauge (\ref{cf}), 
we can drop second or higher order terms in  $g^{++}$ and $g^{--}$ and
the scalar curvature $R$ and de Alembertian $\Box$ have the 
following forms:
\bea
\label{vary}
R&=& 8\e^{-2\rho}\partial_+\partial_-\rho +4Q \ ,\nn
Q&\equiv&g^{--}\left( - \partial_-^2\rho 
-2 (\partial_- \rho)^2 \right)
-{1 \over 4}\partial_-^2 g^{--} - {3 \over 2}
\partial_- g^{--} \partial_-\rho \nn
&& + g^{++}\left( - \partial_+^2\rho 
-2 (\partial_+ \rho)^2 \right)
-{1 \over 4}\partial_+^2 g^{++} - {3 \over 2}
\partial_+ g^{++} \partial_+\rho \ ,\nn
\Box&=&-4\e^{-2\rho}\partial_+\partial_-
+\e^{-2\rho}\partial_+(g^{++}\e^{2\rho}\partial_+)
+\e^{-2\rho}\partial_-(g^{--}\e^{2\rho}\partial_-)\ .
\eea
In the conformal gauge (\ref{cf}), the action $\Gamma_{ind}$ 
(\ref{Gindrd}) is reduced into 
\bea
\label{Gindcf}
{\Gamma_{ind} \over 4\pi}&=&b \int d^2x \e^{2\rho}
\left({32 \over 3}\e^{-4\rho}(\partial_+\partial_-\rho)^2 
+ {16 \over 3} \e^{-2\rho}\partial_+\partial_-\rho 
+ {2 \over 3} \right)\sigma \nn
&& +b' \int d^2x \e^{2\rho}\left\{\sigma\left[
16\e^{-2\rho}\partial_+\partial_-(\e^{-2\rho}\partial_+
\partial_-\sigma) 
\right.\right. \nn
&& + \e^{-2\rho}\left({64 \over 3}\e^{-2\rho}
\partial_+\partial_-\rho 
+ {16 \over 3}\right)\partial_+\partial_-\sigma \nn
&& \left. + {8 \over 3}\e^{-2\rho}\left(\partial_+(\e^{-2\rho}
\partial_+\partial_-\rho)\partial_-\sigma 
+ \partial_-(\e^{-2\rho}
\partial_+\partial_-\rho)\partial_+\sigma \right)\right] \nn
&& \left. + \left(16\e^{-2\rho} \partial_+\partial_-\rho
+ {32 \over 3}\e^{-2\rho} \partial_+\partial_-
(\e^{-2\rho} \partial_+\partial_-\rho)\right)\sigma \right\} \nn
&& -\left\{b''+ {2 \over 3}(b+b')\right\} \int d^2x \e^{2\rho} \nn
&& \times \left\{ 2(\partial_+\partial_-\sigma+ \partial_+\sigma
\partial_-\sigma)\e^{-2\rho}(8\e^{-2\rho}
\partial_+\partial_-\rho + 2) \right. \nn
&& \left. +24 \e^{-4\rho}(\partial_+\partial_-\sigma
+ \partial_+\sigma\partial_-\sigma)^2\right]\ .
\eea
and the variations of $\Gamma_{ind}$ with respect to $g^{\pm\pm}$ are 
given by
\bea
\label{cons1}
{\delta \over \delta g^{\pm\pm}}\left( {\Gamma_{ind} \over 4\pi}\right)
&=& b'\left[ 8\e^{2\rho}\partial_\pm \sigma \partial_\pm 
\left(\e^{-2\rho}\partial_+\partial_- \sigma \right)
-8\sigma\partial_\pm^2\sigma \partial_+\partial_- \rho \right. \nn
&& + {2 \over 3}\e^{2\rho}\partial_\pm\sigma 
\partial_\pm\{(8\e^{-2\rho}\partial_+\partial_- \rho + 2)\sigma \} \nn
&& + {8 \over 3}\e^{2\rho}\sigma 
\partial_\pm(8\e^{-2\rho}\partial_+\partial_- \rho 
+ 2)\partial_\pm \sigma  \nn
&& \left. + {8 \over 3}\e^{2\rho}
\partial_\pm(8\e^{-2\rho}\partial_+\partial_- \rho 
+ 2)\partial_\pm \sigma  \right] \nn
&& -\left\{b''+{2 \over 3}(b+b')\right\}\left[
4e^{2\rho}\partial_\pm\sigma 
\partial_\pm(8\e^{-2\rho}\partial_+\partial_- \rho ) \right.
\nn
&& - 4 (\partial_\pm \sigma)^2\partial_+\partial_-\rho 
+ 12 e^{2\rho}\partial_\pm\sigma \partial_\pm
\{\e^{-2\rho}(\partial_+\partial_-\sigma 
+ \partial_+ \sigma \partial_-\sigma)\} \nn
&& \left. -12 (\partial_+\partial_-\sigma 
+ \partial_+ \sigma \partial_-\sigma)(\partial_\pm\sigma)^2 \right] \nn
&& + \left\{\left(- \partial_\pm^2\rho - 2(\partial_\pm\rho)^2
\right) -{1 \over 4} \partial_\pm^2 
+ {3 \over 2}\partial_\pm^2\rho 
+ {3 \over 2} \partial_\pm\rho \partial_\pm \right\} \nn
&& \times \Biggl[ b \left({32 \over 3} \partial_+\partial_-\rho 
+ {8 \over 3}\e^{2\rho}\right)\sigma 
+ {16 \over 3}b'(-\sigma\partial_+\partial_-\sigma 
+ \partial_+\sigma\partial_-\sigma) \nn
&& \left. -\left\{b''+{2 \over 3}(b+b')\right\}
(\partial_+\partial_-\sigma + \partial_+\sigma\partial_-\sigma) \right]
\eea
Usually the equation given by $g^{++}$ or $g^{--}$ can be regarded as 
the constraint equation with respect to the initial or boundary values.
The equations obtained here are, however, combinations of the constraint and 
$\sigma$-equation of the motion since 
the tensor $g_{\mu\nu}$ under consideration is the product of the 
original metric tensor in (\ref{OIX}) and the $\sigma$-function 
$\e^{-2\sigma}$.

Polyakov's anomaly induced action (\ref{OXI})
has the following form in the conformal gauge (\ref{cf}) as usual
\be
\label{Plycf}
\Gamma^{(0)}=-{N \over 12\pi}\int d^2 x \rho\partial_+ \partial_- \rho 
\ee
and the variations with respect to $g^{\pm\pm}$ are given by
\be
\label{Plycon}
{\delta \Gamma^{(0)} \over \delta g^{\pm\pm}}
={N \over 48\pi}\e^{2\rho}\left(\partial_\pm^2 \rho
- (\partial_\pm \rho)^2\right) + t^\pm(x^\pm)\e^{2\rho}\ .
\ee
Here $t^\pm(x^\pm)$ is a function determined by the boundary condition.
The ``potential'' term $\Gamma^{(V)}$ in (\ref{OXII})
has the following form in the conformal gauge (\ref{cf})
\be
\label{Vaccf}
\Gamma^{(V)}={N \over 96\pi}\int d^2x \e^{2\rho}\left(
8\e^{-2\rho}\partial_+\partial_-\rho +2\right)
\left(1-\ln {8\e^{-2\rho}\partial_+\partial_-\rho 
+2 \over 6\mu^2}\right)
\ee
and the variations with respect to $g^{\pm\pm}$ are given by
\bea
\label{cons3}
{\delta \Gamma^{(V)} \over \delta g^{\pm\pm}}
&=&-{N \over 24\pi}
\left\{\left(- \partial_\pm^2\rho - 2(\partial_\pm\rho)^2
\right) -{1 \over 4} \partial_\pm^2 + {3 \over 2}\partial_\pm^2\rho 
+ {3 \over 2} \partial_\pm\rho \partial_\pm \right\} \nn
&& \times \e^{2\rho}\ln {8\e^{-2\rho}\partial_+\partial_-\rho 
+2 \over 6\mu^2}\ .
\eea
The variations with respect to $\rho$ are given by 
\bea
\label{rhov}
{\delta \over \delta\rho}\left({\Gamma_{ind} \over 4\pi}\right)
&=&b \left\{ -{64 \over 3}\e^{-2\rho}(\partial_+\partial_- \rho)^2\sigma 
+ {64 \over 3}\partial_+\partial_-\left(\sigma\e^{-2\rho}
\partial_+\partial_-\rho\right) \right. \nn
&& \left. + {16 \over 3}\partial_+\partial_-\sigma
+ {4 \over 3}\e^{2\rho}\sigma \right\} \nn
&& + b' \left\{ -32(\partial_+\partial_-\sigma)^2\e^{-2\rho}
-{128 \over 3}\e^{-2\rho} \partial_+\partial_-\rho 
(\sigma\partial_+\partial_-\sigma) \right. \nn
&& + {64 \over 3}\partial_+\partial_-\left(\e^{-2\rho}
\sigma\partial_+\partial_-\sigma\right) \nn
&& - {16 \over 3}\left\{-2\partial_+\sigma\partial_-\sigma
\e^{-2\rho}\partial_+\partial_-\rho 
+ \partial_+\partial_-\left(\partial_+\sigma\partial_-\sigma
\e^{-2\rho}\right)\right\} \nn
&& + 16\partial_+\partial_-\sigma \nn
&& + {32 \over 3}\left\{-2\partial_+\partial_-\sigma
\e^{-2\rho}\partial_+\partial_-\rho 
+ \partial_+\partial_-\left(\partial_+\partial_-\sigma
\e^{-2\rho}\right)\right\} \nn
&& -\left\{b''+ {2 \over 3}(b+b')\right\}\left[
16 \partial_+\partial_- \left\{ \e^{-2\rho} (\partial_+\partial_-\sigma
+\partial_+\sigma\partial_-\sigma)\right\} \right. \nn
&& \left. -48\e^{-2\rho}(\partial_+\partial_-\sigma
+\partial_+\sigma\partial_-\sigma)^2\right] \\
\label{rho0}
{\delta \Gamma^{(0)} \over \delta \rho}&=&
- {N \over 6\pi}\partial_+\partial_-\rho \\
\label{rhoV}
{\delta \Gamma^{(V)} \over \delta \rho}&=&
+ {N \over 48\pi}\e^{2\rho}\left(
8\e^{-2\rho}\partial_+\partial_-\rho +2 \right)
\left(1-\ln {8\e^{-2\rho}\partial_+\partial_-\rho
+2 \over 6\mu^2}\right) \nn
&& -{N \over 12\pi}\partial_+\partial_-\left(
\ln {8\e^{-2\rho}\partial_+\partial_-\rho +2 \over 6\mu^2}\right) \nn
&& + {N \over 6\pi}\e^{2\rho}\partial_+\partial_-\rho 
\ln {8\e^{-2\rho}\partial_+\partial_-\rho +2 \over 6\mu^2}\ .
\eea
The variations with respect to $\sigma$  may be found as
%%%%%%%%%%%%%%%
\bea
\label{sigmav}
{\delta \over \delta \sigma} \left({\Gamma_{ind} \over 4\pi}\right)
&=&b \e^{2\rho}\left({32 \over 3}\e^{-4\rho}
(\partial_+\partial_-\rho)^2 + {32 \over 3}\e^{-2\rho}
\partial_+\partial_-\rho + {2 \over 3}\right) \nn
&& + b'\left[ 32 \partial_+\partial_-(
\e^{-2\rho}\partial_+\partial_-\sigma) \right. \nn
&& +{64 \over 3}\left(\e^{-2\rho}
\partial_+\partial_-\rho\partial_+\partial_-\sigma
+ \partial_+\partial_-(\sigma\e^{-2\rho}
\partial_+\partial_-\rho)\right) \nn
&& +{32 \over 3}\partial_+\partial_-\sigma \nn
&& + {16 \over 3}\left\{\partial_+(\e^{-2\rho}
\partial_+\partial_-\rho)\partial_-\sigma +\partial_-(\e^{-2\rho}
\partial_+\partial_-\rho)\partial_+\sigma\right\} \nn
&& \left. + \left( 16 \partial_+\partial_-\rho
+ {32 \over 3}\partial_+\partial_-(\e^{-2\rho}
\partial_+\partial_-\rho)\right)\right]\nn
&& -\left\{b''+ {2 \over 3}(b+b')\right\}\left[
2\partial_+\partial_-(8\e^{-2\rho}
\partial_+\partial_-\rho +2) \right. \nn
&& -2\left\{\partial_+(8\e^{-2\rho}
\partial_+\partial_-\rho +2)\partial_-\sigma +\partial_-(8\e^{-2\rho}
\partial_+\partial_-\rho +2)\partial_+\sigma\right\} \nn
&& +48\partial_+\partial_-\left\{ \e^{-2\rho}
(\partial_+\partial_-\sigma
+\partial_+\sigma\partial_-\sigma)\right\} \nn
&& -48\partial_+\left\{ \e^{-2\rho}\partial_-\sigma
(\partial_+\partial_-\sigma
+\partial_+\sigma\partial_-\sigma)\right\} \nn
&& \left. -48\partial_-\left\{ \e^{-2\rho}\partial_+\sigma
(\partial_+\partial_-\sigma
+\partial_+\sigma\partial_-\sigma)\right\} \right] \\
\label{sgmvo}
{\delta \Gamma^{(0)} \over \delta \sigma}&=&
{\delta \Gamma^{(V)} \over \delta \sigma}=0
\eea
It often happens that we can drop the terms linear to $\sigma$ 
in (\ref{Gindrd}). In particular, one can redefine the corresponding 
source term as it is in the case of IR sector of 4D QG \cite{AMO}. In that case, Eqs. 
(\ref{cons1}), (\ref{rhov}) and (\ref{sigmav}) are reduced into
\bea
\label{cons1b}
{\delta \over \delta g^{\pm\pm}}\left( {\Gamma_{ind} \over 4\pi}\right)
&=&b'\left[ 8\e^{2\rho}\partial_\pm \sigma \partial_\pm 
\left(\e^{-2\rho}\partial_+\partial_- \sigma \right)
-8\sigma\partial_\pm^2\sigma \partial_+\partial_- \rho \right. \nn
&& + {2 \over 3}\e^{2\rho}\partial_\pm\sigma 
\partial_\pm\{(8\e^{-2\rho}\partial_+\partial_- \rho + 2)\sigma \} \nn
&& \left. + {8 \over 3}\e^{2\rho}\sigma 
\partial_\pm(8\e^{-2\rho}\partial_+\partial_- \rho 
+ 2)\partial_\pm \sigma  \right] \nn
&& -\left\{b''+{2 \over 3}(b+b')\right\}\left[
4e^{2\rho}\partial_\pm\sigma 
\partial_\pm(8\e^{-2\rho}\partial_+\partial_- \rho ) \right. \nn
&& - 4 (\partial_\pm \sigma)^2\partial_+\partial_-\rho 
+ 12 e^{2\rho}\partial_\pm\sigma \partial_\pm
\{\e^{-2\rho}(\partial_+\partial_-\sigma 
+ \partial_+ \sigma \partial_-\sigma)\} \nn
&& \left.-12 (\partial_+\partial_-\sigma 
+ \partial_+ \sigma \partial_-\sigma)(\partial_\pm\sigma)^2 \right] \nn
&& + \left\{\left(- \partial_\pm^2\rho - 2(\partial_\pm\rho)^2
\right) -{1 \over 4} \partial_\pm^2 + {3 \over 2}\partial_\pm^2\rho 
+ {3 \over 2} \partial_\pm\rho \partial_\pm \right\} \nn
&& \times \Biggl[ {16 \over 3}b'(-\sigma\partial_+\partial_-\sigma 
+ \partial_+\sigma\partial_-\sigma) \nn
&& \left. -\left\{b''+{2 \over 3}(b+b')\right\} (\partial_+\partial_-\sigma 
+ \partial_+\sigma\partial_-\sigma) \right]\ ,\\
\label{rhovb}
{\delta \over \delta\rho}\left({\Gamma_{ind} \over 4\pi}\right)
&=& b' \left\{ -32(\partial_+\partial_-\sigma)^2\e^{-2\rho}
-{128 \over 3}\e^{-2\rho} \partial_+\partial_-\rho 
(\sigma\partial_+\partial_-\sigma) \right. \nn
&& + {64 \over 3}\partial_+\partial_-\left(\e^{-2\rho}
\sigma\partial_+\partial_-\sigma\right) \nn
&& - {16 \over 3}\left\{-2\partial_+\sigma\partial_-\sigma
\e^{-2\rho}\partial_+\partial_-\rho 
+ \partial_+\partial_-\left(\partial_+\sigma\partial_-\sigma
\e^{-2\rho}\right)\right\} \nn
&& -\left\{b''+ {2 \over 3}(b+b')\right\}\left[
16 \partial_+\partial_- \left\{ \e^{-2\rho} (\partial_+\partial_-\sigma
+\partial_+\sigma\partial_-\sigma)\right\} \right. \nn
&& \left. -48\e^{-2\rho}(\partial_+\partial_-\sigma
+\partial_+\sigma\partial_-\sigma)^2\right] \\
\label{sigmavb}
{\delta \over \delta \sigma} \left({\Gamma_{ind} \over 4\pi}\right)
&=& b'\left[ 32 \partial_+\partial_-(
\e^{-2\rho}\partial_+\partial_-\sigma) \right. \nn
&& +{64 \over 3}\left(\e^{-2\rho}
\partial_+\partial_-\rho\partial_+\partial_-\sigma
+ \partial_+\partial_-(\sigma\e^{-2\rho}
\partial_+\partial_-\rho)\right) \nn
&& +{32 \over 3}\partial_+\partial_-\sigma \nn
&& \left. + {16 \over 3}\left\{\partial_+(\e^{-2\rho}
\partial_+\partial_-\rho)\partial_-\sigma +\partial_-(\e^{-2\rho}
\partial_+\partial_-\rho)\partial_+\sigma\right\} \right]\nn
&& -\left\{b''+ {2 \over 3}(b+b')\right\}\left[
2\partial_+\partial_-(8\e^{-2\rho}
\partial_+\partial_-\rho +2) \right. \nn
&& -2\left\{\partial_+(8\e^{-2\rho}
\partial_+\partial_-\rho +2)\partial_-\sigma +\partial_-(8\e^{-2\rho}
\partial_+\partial_-\rho +2)\partial_+\sigma\right\} \nn
&& +48\partial_+\partial_-\left\{ \e^{-2\rho}
(\partial_+\partial_-\sigma +\partial_+\sigma\partial_-\sigma)\right\} \nn
&& -48\partial_+\left\{ \e^{-2\rho}\partial_-\sigma
(\partial_+\partial_-\sigma
+\partial_+\sigma\partial_-\sigma)\right\} \nn
&& \left. -48\partial_-\left\{ \e^{-2\rho}\partial_+\sigma
(\partial_+\partial_-\sigma
+\partial_+\sigma\partial_-\sigma)\right\} \right] 
\eea
The classical Einstein action
\be
\label{cS}
S=-{1 \over 16\pi G}\int d^4x \sqrt{-g_{(4)}}
\left\{R^{(4)} -2\Lambda\right\} 
\ee
is reduced into 2 dimensional one in the conformal gauge 
(\ref{cf}) as follows;
\be
\label{cSred}
{S \over 4\pi}=-{1 \over 16\pi G}\int d^2x 
\left[ 4\e^{2\sigma}\partial_+\partial_-(\rho + \sigma)
+2\e^{2\rho+2\sigma}-\e^{2\rho+4\sigma}\Lambda 
-4 \e^{2\sigma}\partial_+\sigma\partial_-\sigma \right],
\ee
and we obtain
\bea
\label{cScons}
{\delta \over \delta g^{\pm\pm}}
\left({S \over 4\pi}\right)&=&-{1 \over 16\pi G}\e^{2\rho+2\sigma}
\left[(\partial_\pm\sigma)^2 -\partial_\pm^2\sigma
+2\partial_\pm\sigma\partial_\pm\rho\right] \\
\label{cSsigma}
{\delta \over \delta \sigma} 
\left({S \over 4\pi}\right)&=&-{1 \over 16\pi G}
\Biggl[8\e^{2\sigma}\left\{3\partial_+\partial_- \sigma
+3\partial_+\sigma\partial_-\sigma + \partial_+\partial_-\rho \right\} \nn
&& -4\e^{2\rho+4\sigma}\Lambda + 4\e^{2\rho+2\sigma} \Biggr] \\
\label{cSrho}
{\delta \over \delta \rho} \left({S \over 4\pi}\right)&=&-{1 \over 16\pi G}
\left[4\partial_+\partial_-\e^{2\sigma}
-2\e^{2\rho+4\sigma}\Lambda + 4\e^{2\rho+2\sigma}\right]\ .
\eea
Above equations give the complete system of 
equations of motion for the system under discussion.

\section{Anti-evaporation of Nariai black holes}

Following the work by Bousso-Hawking \cite{BH} we consider the 
Schwarzschild-de Sitter family of black holes, or more exactly its 
nearly degenerated case, so-called Nariai solution \cite{N}. 
We are now interested in the instability of the Nariai solution.
The Schwarzschild type black hole solution in the de Sitter space has 
two horizons, one is the usual event horizon and another is the 
cosmological horizon, which is proper one  in the de Sitter space. 
The Nariai solution is given by a limit of the 
Schwarzschild-de Sitter black hole where two horizons coincides with 
each other. In the limit, the two horizons have the 
same temperature and they are in the thermal equilibrium. 
Near the limit, however, the temperature of the event horizon is 
higher than that of the cosmological one and we can expect that there 
would be a thermal flow from the event horizon to the cosmological one.
Therefore the system would become instable and 
the black hole would evaporate. We also have to note that above 
cosmological black holes may naturally appear only through 
quantum pair creation \cite{GP} which may occur in the 
inflationary universe \cite{BH2}.

In the Nariai limit, the space-time has the topology of 
$S^1\times S^2$ and the metric is given by
\be
\label{Nsol}
ds^2={1 \over \Lambda}\left( \sin^2 \chi d\psi^2 - d\chi^2 - d\Omega
\right)\ .
\ee
Here the coordinate $\chi$ has a period $\pi$.
If we change the coordinates variables by
\bea
\label{cv}
r&=&\ln\,\tan{\chi \over 2} \nn
t&=&{\psi \over 4}\ ,
\eea
we obtain 
\be
\label{Nsol2}
ds^2= {1 \over \Lambda\cosh^2 x}\left(-dt^2 + dr^2\right) 
+ {1 \over \Lambda}d\Omega\ .
\ee
This form would correspond to the conformal gauge in two dimensions.
Note that the transformation (\ref{cv}) has one to one correspondence 
between $(\psi,\chi)$ and $(t,r)$ if we restrict $\chi$ by 
$0\leq \chi <\pi$ ($r$ runs from $-\infty$ to $+\infty$). 

Now we solve the equations of motion. Since the Nariai solution 
is characterized by the constant $\phi$(or $\sigma$),
we now assume that $\sigma$ is a constant even when including quantum 
correction;
\be
\label{consigma}
\sigma=\sigma_0\ \ \mbox{(constant)}\ .
\ee
We also first consider static solutions and replace 
$\partial_{\pm}$ by $\pm{1 \over 2}\partial_r$.
Then the total constraint equation obtained by (\ref{Plycon}), 
(\ref{cons3}), (\ref{cons1b}) and (\ref{cScons}) becomes 
\bea
\label{stcons}
0&=&-{N \over 192\pi}\e^{2\rho}\left(\partial_r^2\rho 
-(\partial_r\rho)^2\right) + t_0 \e^{2\rho} \nn
&& +{N \over 96\pi}
\left\{ \partial_r^2\rho + 2(\partial_r\rho)^2
+ {1 \over 4} \partial_r^2 - {3 \over 2}\partial_r^2\rho 
- {3 \over 2} \partial_r \rho \partial_r \right\} \nn
&& \times \e^{2\rho}\ln {-2\e^{-2\rho}\partial_r^2\rho 
+2 \over 6\mu^2}\ .
\eea
The equations of motions given by (\ref{rho0}), (\ref{rhoV}), 
(\ref{rhovb}) and (\ref{cSrho}) (using the variation of $\rho$)
(\ref{sgmvo}), (\ref{sigmavb}) and (\ref{cSsigma}) (using the variation 
of $\sigma$) have the following forms:
\bea
\label{strho}
0&=&-{1 \over 4\pi G}\left(-2\Lambda\e^{2\rho+4\sigma_0}
+4\e^{2\rho+2\sigma_0}\right) \nn
&& + {16\pi \over 3}b\sigma_0\left\{ -\e^{-2\rho}(\partial_r^2\rho)^2 
+ \partial_r^2(\e^{-2\rho}\partial_r^2\rho) + \e^{2\rho} \right\} \nn
&& - {N \over 24\pi}\e^{2\rho}
\left(1-\ln {-2\e^{-2\rho}\partial_r^2\rho +2 \over 6\mu^2}\right) \nn
&& -{N \over 48\pi}\partial_r^2\left( \ln {-2\e^{-2\rho}\partial_r^2\rho
+2 \over 6\mu^2}\right) \\
\label{stsigma}
0&=&-{1 \over 4\pi G}\left( -2\e^{2\sigma_0}\partial_r^2\rho 
-4\Lambda\e^{2\rho+4\sigma_0} + 4\e^{2\rho+2\sigma_0} \right) \nn
&& + 16\pi\left\{b''+{2 \over 3}(b+b')\right\}
\partial_r^2\left(\e^{-2\rho}\partial_r^2\rho\right)\ .
\eea
 Assuming a solution is given by a constant 2d scalar 
curvature
\be
\label{conscur}
R=-2\e^{-2\rho}\partial_r^2\rho=R_0\ \ 
(\mbox{constant})
\ee
and substituting (\ref{conscur}) into (\ref{strho}) and 
(\ref{stsigma}), we find that Eqs. (\ref{strho}) and 
(\ref{stsigma}) can be satisfied if $\sigma_0$ and $R_0$ 
satisfy the following two algebraic equations
\bea
\label{aleq1}
0&=&R_0 - 4\Lambda \e^{2\sigma_0} + 4 \\
\label{aleq2}
0&=& -{1 \over 2\pi G}\left(-\Lambda\e^{4\sigma_0} +2\e^{2\sigma_0}\right) 
+ {4\pi b \over 3}\left( -R_0^2 + 4\right)\sigma_0 \nn
&& - {N \over 24\pi}R_0 \left(1 - \ln {R_0 + 2 \over 6\mu^2 }\right)\ .
\eea
The above equation can be solved with respect to $\sigma_0$ and $R_0$ 
in general although it is difficult to get the explicit expression. 
A special case is given by dropping the term corresponding 
to the terms linear to $\sigma$ in (\ref{Gindrd}) 
(i.e., by dropping the second term proportional 
to $b$ in (\ref{aleq2})) and by choosing the renormalization 
scale $\mu$ to be 
\be
\label{renscl}
\mu^2={R_0 + 2 \over 6}\ .
\ee
In this case, the solution of (\ref{aleq1}) and (\ref{aleq2}) is given by 
\bea
\label{solaleq}
R_0&=&\pm 4 \sqrt{1 + {NG\Lambda \over 12}} \nn
\e^{2\sigma}&=&{1 \over \Lambda}\left( 1 \pm 
\sqrt{1 + {NG\Lambda \over 12}} \right)\ .
\eea
If we choose $+$ sign in front of the root in the above 
equations, the solution reduces to the classical 
solution in the classical limit of $N\rightarrow 0$.

Equation (\ref{conscur}) can be integrated to be
\be
\label{intcur}
(\partial_r\rho)^2=-{R_0 \over 2}\e^{2\rho} + C\ .
\ee
Here $C$ is a constant of integration.  Substituting
(\ref{conscur}) and (\ref{intcur}) into (\ref{stcons}), we find 
Eq. (\ref{stcons}) is satisfied if and only if 
\be
\label{tC}
t_0=-{N \over 192\pi}C\ .
\ee
Eq.(\ref{intcur}) can be integrated to become
\be
\label{rho}
\e^{2\rho}={2C \over R_0}\cdot{1 \over \cosh^2 
\left(r\sqrt{C}\right)} .
\ee

We now consider the perturbation around the Nariai 
type solution (\ref{consigma}) and (\ref{rho});
\bea
\label{pert}
\rho&=&\rho_0 + \epsilon R(t,r) \nn
\sigma&=&\sigma_0 + \epsilon S(t,r) \nn
t_0&=&{N \over 192\pi}C + \epsilon T^\pm\ .
\eea
Here $\epsilon$ is a infinitesimaly small parameter and 
$\rho_0$ is a solution given by (\ref{rho}):
\be
\label{rho0s}
\e^{2\rho_0}={2C \over R_0}\cdot{1 \over \cosh^2 
\left(r\sqrt{C}\right)} .
\ee
Then we obtain the following equations
\bea
\label{pertcons}
0&=&-8\pi{b'\sigma_0 C \over \cosh^2\left(r\sqrt{C}\right)}
\partial_\pm^2S \nn
&& + 4\pi \left(-{16 \over 3}b'\sigma_0 - {2 \over 3}(b+b')\right)
\left({3 \over 8}{1 \over \cosh^2\left(r\sqrt{C}\right)}
-{1 \over 2}\right)C\partial_+\partial_-S \nn
&& + 4\pi \left(-{16 \over 3}b'\sigma_0 - {2 \over 3}(b+b')\right)
\left(-{1 \over 4}\partial_\pm^2\partial_+\partial_-S
\mp{3 \over 4}\sqrt{C}\tanh \left(r\sqrt{C}\right)
\partial_\pm\partial_+\partial_-S\right) \nn
&& - {N \over 24\pi}{C \over R_0}\left(-{C \over 2}R + \partial_\pm^2 R 
\pm \sqrt{C}\tanh\left(r\sqrt{C}\right) \partial_\pm R \right) \nn
&& -{2C \over R_0}{1 \over \cosh^2\left(r\sqrt{C}\right)}
\left({NC \over 96\pi} + T^\pm\right) \nn
&& + {N \over 24\pi}\left[{C \over 2R_0}\partial_\pm^2 R 
{1 \over \cosh^2\left(r\sqrt{C}\right)}\ln{R_0 + 2 \over 6\mu^2} \right. \nn
&& -{4C \over R_0}{1 \over \cosh^2\left(r\sqrt{C}\right)}
\left(-{1 \over 4}\partial_\pm^2R 
\mp {\sqrt{C} \over 2}\tanh \left(r\sqrt{C}\right)
\partial_\pm R \right)  \nn
&& +{8C \over R_0}\left\{ \left({3 \over 8}
{1 \over \cosh^2\left(r\sqrt{C}\right)}-{1 \over 2}\right)
\partial_+\partial_-R \right. \nn
&& \left.\left. - {1 \over 4}\partial_\pm^2\partial_+\partial_-R 
\mp {3\sqrt{C} \over 4}\tanh\left(r\sqrt{C}\right)
\partial_\pm\partial_+\partial_-R\right\} \right] \nn
&& - {1 \over 16\pi G}{2C \over R_0}
{\e^{2\sigma_0} \over \cosh^2\left(r\sqrt{C}\right)}
\left(-\partial_\pm^2 S  \mp\sqrt{C}\tanh\left(r\sqrt{C}\right)
\partial_\pm S \right) \\
\label{pertrho}
0&=&4\pi b'\left\{-{16 \over 3}R_0\sigma_0\partial_+
\partial_-S +{32 \over 3}{R_0\sigma_0 \over C}\partial_+
\partial_-\left(\cosh^2\left(r\sqrt{C}\right)\partial_+
\partial_-S\right) \right\}\nn
&& -4\pi(b+b'){16 \over 3}{R_0 \over C}\partial_+
\partial_-\left(\cosh^2\left(r\sqrt{C}\right)\partial_+
\partial_-S\right)  \nn
&& -{N \over 12\pi }{C \over R_0}{1 \over \cosh^2\left(r\sqrt{C}\right)} \nn
&& \times\left\{2R\left(1- \ln{R_0 + 2 \over 6\mu^2}\right) 
-{1 \over R_0 + 2}\left(-2R_0R + {4R_0 \over C} 
\cosh^2\left(r\sqrt{C}\right)\partial_+\partial_- R\right)\right\} \nn
&& +{N \over 12\pi}{1 \over R_0 + 2}\left\{
-2R_0\partial_+\partial_-R  + {4R_0 \over C}
\partial_+\partial_-\left( \cosh^2\left(r\sqrt{C}\right)
\partial_+\partial_-R \right)\right\} \nn 
&& - {1 \over 16\pi G}\e^{2\sigma_0}\left[ 8\partial_+\partial_-S 
- {8\Lambda C \over R_0}{1 \over \cosh^2\left(r\sqrt{C}\right)}
 \e^{2\sigma_0}(R+2S) \right. \nn
&& \left. + {16 C \over R_0} 
{1 \over \cosh^2\left(r\sqrt{C}\right)}(R+S) \right] \\
\label{pertsigma}
0&=&4\pi b'\left\{{128 \over 3}R_0 \partial_+
\partial_-S +16{R_0 \over C}\partial_+
\partial_-\left(\cosh^2\left(r\sqrt{C}\right)\partial_+
\partial_-S\right) \right. \nn
&& \left. - {16 \over 3}\sigma_0R_0 \partial_+
\partial_-R +  {32 \over 3}{\sigma_0 R_0 \over C}\partial_+
\partial_-\left( \cosh^2\left(r\sqrt{C}\right)
\partial_+\partial_-R \right) \right\} \nn
&& - 4\pi\times 16(b+b'){R_0 \over C}\partial_+
\partial_-\left(\cosh^2\left(r\sqrt{C}\right)\partial_+
\partial_-S\right) \nn
&& - {1 \over 16\pi G}\e^{2\sigma_0}\left\{8\left(
3\partial_+\partial_-S + 
\partial_+\partial_-R + {C \over 2}{1 \over 
\cosh^2\left(r\sqrt{C}\right)}S\right) \right. \nn
&& \left. -{16\Lambda C \over R_0} {\e^{2\sigma_0} \over 
\cosh^2\left(r\sqrt{C}\right)}(R+2S)
+ {16 C \over R_0} {1 \over \cosh^2\left(r\sqrt{C}\right)}(R+S)
\right\}
\eea
When deriving the above equations, we have dropped 
the terms linear to $\sigma$ in (\ref{Gindrd}), for 
simplicity. 

The equations (\ref{pertrho}) and (\ref{pertsigma}) can be 
solved by assuming $R$ and $S$ are given by
\bea
\label{RS}
R(t,r)&=&P\e^{\pm t\alpha\sqrt{C}}\cosh^\alpha r\sqrt{C} \nn
S(t,r)&=&Q\e^{\pm t\alpha\sqrt{C}}\cosh^\alpha r\sqrt{C} \ .
\eea
Here $P$, $Q$ and $\alpha$ are constants. 
We should note that $\alpha<0$ if we require $R$ and $S$ 
are finite in the limit of $r\rightarrow \pm\infty$.
Then we obtain the following equations
\bea
\label{eigenRS}
\cosh^2\left(r\sqrt{C}\right)\partial_+\partial_- R &=& A R \nn
\cosh^2\left(r\sqrt{C}\right)\partial_+\partial_- S &=& A S \nn
A&\equiv& {\alpha (\alpha - 1) C \over 4}\ .
\eea
Note that there is one to one correspondence between $A$ and 
$\alpha$ if we restrict $A>0$ and $\alpha<0$. 
Using (\ref{eigenRS}), we can rewrite (\ref{pertrho}) and 
(\ref{pertsigma}) (using (\ref{aleq1})) as follows:
\bea
\label{perte}
0&=& \left\{ {64\pi b' \sigma_0 R_0 \over 3}\left( - A
+ 2 {A^2 \over C}\right) - {64\pi (b+b') R_0 \over 3}{A^2 \over C}
- {1 \over 4\pi G}\e^{2\sigma_0}\left(2A - C \right) \right\}Q \nn
&& + \left\{ -{N \over 6\pi}{C \over R_0}
\left(1 + \ln {R_0 +2 \over 6\mu^2}\right) 
+{N \over 6\pi}{C \over R_0 + 2}
\left( - 1 + {2 \over C}A \right) \right. \nn
&& \left. + {N \over 6\pi}{R_0 \over R_0 + 2}
\left( - A + {2 \over C}A^2 \right)
- {C \over 8\pi G}\e^{2\sigma_0}\left( -1 + 
{4 \over R_0}\right)\right\}P \nn
0&=&\left\{ 64\pi b' R_0 \left( {8 \over 3}A + {1 \over C}A^2 \right) 
- 64\pi (b+b'){R_0 \over C}A^2 \right. \nn
&& \left. - {\e^{2\sigma_0} \over 4\pi G}
\left(6A - C - {4C \over R_0} \right) \right\}Q \nn
&& + \left\{ {64\pi b' \sigma_0 R_0 \over 3}\left( -A 
+ {2 \over C}A^2 \right) - {\e^{2\sigma_0} \over 4\pi G}(2A-C) \right\}P\ .
\eea
In order that the above two algebraic equations have a 
non trivial solution for $P$ and $Q$, $A$ should satisfy 
the following equation
\bea
\label{egneq}
0&=&\left\{ {64\pi b' \sigma_0 R_0 \over 3}\left( - A
+ 2 {A^2 \over C}\right) - {64\pi (b+b') R_0 \over 3}{A^2 \over C}
- {1 \over 4\pi G}\e^{2\sigma_0}\left(2A - C \right) \right\} \nn
&& \times \left\{ {64\pi b' \sigma_0 R_0 \over 3}\left( -A 
+ {2 \over C}A^2 \right) - {\e^{2\sigma_0} \over 4\pi G}(2A-C) \right\} \nn
&& - \left\{ -{N \over 6\pi}{C \over R_0}
\left(1 + \ln {R_0 +2 \over 6\mu^2}\right) 
+{N \over 6\pi}{C \over R_0 + 2}
\left( - 1 + {2 \over C}A \right) \right. \nn
&& \left. + {N \over 6\pi}{R_0 \over R_0 + 2}
\left( - A + {2 \over C}A^2 \right)
- {C \over 8\pi G}\e^{2\sigma_0}\left( -1 + 
{4 \over R_0}\right)\right\} \nn
&& \times \left\{ 64\pi b' R_0 \left( {8 \over 3}A 
+ {1 \over C}A^2 \right) - 64\pi (b+b'){R_0 \over C}A^2 \right. \nn
&& \left. - {\e^{2\sigma_0} \over 4\pi G}
\left(6A - C - {4C \over R_0} \right) \right\} \nn
&\equiv& F(A)\ .
\eea
If the equation (\ref{egneq}) has a positive ($A>0$) solution for $A$, 
i.e., a real negative solution for $\alpha$, there is a solution where 
$S$ increases exponentially in time and the system would be unstable.
It is generally difficult to find the parameter region where $A$ 
has a positive solution but it might not be so difficult to show the 
existence of such a parameter region. 

In order to show the existence, we now consider the solution of 
(\ref{solaleq}) and  $R_0\rightarrow 4$ limit 
($NG\Lambda \rightarrow 0$). Then $F(0)$ is given by 
\be
\label{F0}
F(0)\rightarrow {C^2 \over 4\pi^2G^2 \Lambda^2} >0\ .
\ee
On the other hand, when $A={C \over 2}$
\be
\label{FC2}
F\left(A={C \over 2}\right)\rightarrow -{N \over 24\pi}
\left({91 \pi b \over 2^3\cdot 3^2 \cdot 5 \pi}
+{1 \over 2\pi\Lambda G}\right)C^2 < 0\ .
\ee
Eqs.(\ref{F0}) and (\ref{FC2}) tell that there is a solution of 
the equation $F(A)=0$ when $0<A<{C \over 2}$, i.e., 
\be
\label{alpha}
0>\alpha>-1\ .
\ee
In order to show that the existence of the solution (\ref{alpha}) 
implies the instability of the system, we consider a linear combination of 
(\ref{RS}):
\be
\label{cshS}
S(t,r)=Q\cosh t\alpha\sqrt{C}\cosh^\alpha r\sqrt{C} \ .
\ee
It should be noted that any linear combination 
of two solutions is always a solution since the 
perturbative equations of motion are linear differential 
equations.
The horizon is given by the condition
\be
\label{horizon}
\nabla\sigma\cdot\nabla\sigma=0\ .
\ee
Substituting (\ref{cshS}) into (\ref{horizon}), 
we find the horizon is given by
\be
\label{horizonb}
r=\alpha t\ .
\ee
Therefore on the horizon, we obtain
\be
\label{Shrzn}
S(t,r(t))=Q\cosh^{1+\alpha} t\alpha\sqrt{C} \ .
\ee
It should be noted $1+\alpha>0$ in the solution of (\ref{alpha}), which 
implies the system is unstable.
Since the radius of the horison $r_h$ is given by
\be
\label{hrds}
r_h=\e^\sigma =\e^{\sigma_0 + \epsilon S(t,r(t))}\ ,
\ee
the radius increases monotonically in time if 
the initial perturbation at $t=0$ is positive $Q>0$, on the other 
hand, if the initial perturbation is negative $Q<0$, the radius 
shrinks, i.e., the black hole evaporates.

The solution in $0>\alpha>-1$ shows that the system is unstable.
The perturbation, however, becomes stable if there is a solution 
where $\alpha<-1$, i.e., $A>{C \over 2}$. 
In order to show the existence of the solution where $A>{C \over 2}$, 
we consider $F(A)$ in the limit of $A\rightarrow +\infty$. As in 
the previous case, we consider the limit near the classical solution 
$R_0\sim 4$. 
Then we find
\be
\label{Finf}
F(A\rightarrow +\infty)\rightarrow \left(
{2^{18} \over 3^4}\left( b \ln{\Lambda \over 2}\right)^2 
+ {2^9 \pi N b \over 3^2} \right) {A^4 \over C^2}>0
\ee
Combining (\ref{FC2}) and (\ref{Finf}), we find that there is a solution
where $A>{C \over 2}$ i.e., $\alpha<-1$. This implies that the perturbation 
is stable since the perturbation becomes exponentially small in time 
on the horizon (\ref{Shrzn})
\be
\label{Shrzn2}
S(t,r(t))\rightarrow Q\e^{(1+\alpha) t |\alpha|\sqrt{C}} \ .
\ee
Therefore if the initial perturbation is negative $Q<0$ and the 
perturbed 
radius of horizon is smaller than that of the Nariai limit, 
which is the thermal equilibrium, the radius increases in time and 
approaches to the Nariai limit asymptotically, i.e., the black 
hole anti-evaporates as observed by Bousso and Hawking \cite{BH}.
It might be interesting that there are both of stable and unstable 
perturbations in the model discussed here although the treatment of 
the quantum corections is rather different from that of Bousso and 
Hawking. In the model by Bousso and Hawking, it depends on the initial 
conditions if the perturbation is stable or unstable, i.e., the perturbation 
is stable and the black hole anti-evaporates if its initial value does 
not vanish but its time derivative vanishes, on the other hand, the 
perturbation is unstable if its initial value vanishes but its time 
derivative does not vanish. In the model discussed in this paper, however, 
the initial conditions for the value and the time derivative of 
the stable and unstable perturbations are completely the  
same, i.e., the initial value does not vanish but the initial time derivative 
vanishes for both cases. The reason why these initial conditions lead to the 
physically different results could be because the effective Lagrangian
contains higher derivatives and the initial value and the initial 
time derivative are not sufficient as a set of the initial conditions 
and we need the initial conditions for the higher derivatives to determine 
the dynamics.
%%%%% 

\section{Discussion}

In summary, we demonstrated that in the model under discussion 
black holes may evaporate or anti-evaporate. We have to note that our 
consideration has quite common character. We limited to the case of conformal 
scalars but we could start from the theory of free conformal scalars, spinors 
and vectors. (As is known \cite{BOS} many asymptotically free GUTs in the early 
Universe are also asymptotically conformally invariant and they may be represented 
as collection of free conformal fields). Then, the anomaly induced effective action 
$W$ would have the same form as in section 2 where only coefficients $b$ and $b'$ would be
 changed (but their signs are the same). And in the Eqs.(\ref{OXI}) and 
(\ref{OXII}) again only numerical 
coefficients would be changed due to spinor and vector corrections (in Eq.(\ref{OXII}) also 
signs of some coefficients could be changed). Hence, our result actually corresponds to quantum 
conformal scalar-spinor-vector system.

There are different possibilities to extend our results. From one side, one can study 
the similar, anti-evaporation effect in black holes realised in gravity with torsion. 
The corresponding anomaly induced effective action is given in last ref. of \cite{R}. 
That could help to find finally some phenomenon where torsion effects may be significally strong to prove
(or disprove) the existence of torsion in the Nature.

 From another side, using the effective action of section 2 one can investigate 
the quantum cosmology of Kantowski-Sacks form \cite{KS}. Recently, such cosmology has 
been studied in large $N$ and $s$-waves approximation \cite{KNO}. In the above works it was found that realization of non-singular Universe for 
the models under discussion is problematic. The results of the present work (if exchange 
time and radius) indicate to the possibility of construction of non-singular 
Kantowski-Sacks early Universe.

\noindent
{\bf Acknoweledgments}
 We would like to thank R. Bousso and S. Hawking for useful e-mail discussions.
The work by SDO has been supported partially by Universidad del Valle and 
COLCIENCIAS(COLOMBIA).

\end{document}